\documentclass[english,11pt,oneside]{article}
\pdfoutput=1
\usepackage{amsmath}
\usepackage{amsfonts}
\usepackage{amssymb}
\usepackage{graphicx, rotating}
\usepackage{epstopdf}
\usepackage{epsfig}
\usepackage{latexsym}
\usepackage{multirow}
\usepackage{color}
\usepackage{slashed}
\usepackage{pifont}
\usepackage{soul}
\usepackage{mathtools}
\usepackage[top=2.8 cm, bottom=2.8 cm, left=3 cm, right=3 cm]{geometry}
\usepackage[compat=1.0.0]{tikz-feynman}
\usetikzlibrary{shapes.misc}
\tikzset{cross/.style={cross out, draw=black, minimum size=2*(#1-\pgflinewidth), inner sep=0pt, outer sep=0pt},
	cross/.default={1pt}}
\usepackage{subcaption}
\newcommand{\ba}{\begin{eqnarray}}
\newcommand{\ea}{\end{eqnarray}}
\newcommand{\no}{\nonumber}
\newcommand{\tr}{{\rm Tr}}

\newcommand\blfootnote[1]{%
	\bgroup
	\renewcommand\thefootnote{\fnsymbol{footnote}}%
	\renewcommand\thempfootnote{\fnsymbol{mpfootnote}}%
	\footnotetext[0]{#1}%
	\egroup
}

\usepackage{hyperref}
\hypersetup{
 colorlinks = true,
 linkcolor  = red,
 citecolor={blue}
}

\usepackage{bbold}

\usepackage[nottoc]{tocbibind}

\numberwithin{equation}{section}

\begin{document}

\unitlength = 1mm

\thispagestyle{empty} 
\begin{center}
	\vskip 3.4cm\par
	{\par\centering \textbf{\LARGE Perturbative running of the topological angles}}
	
	\vskip 1.2cm\par
	{\scalebox{.85}{\par\centering \large  
			\sc\hyperlink{AV}{\color{black}Alessandro Valenti}$\,^{a,b}$, \hyperlink{LV}{\color{black}Luca Vecchi}$\,^{b}$}
		{\par\centering \vskip 0.7 cm\par}
		{\sl 
			$^a$~Dipartamento di Fisica e Astronomia ``G.~Galilei", Università di Padova, Italy
		}\\
		{\par\centering \vskip 0.2 cm\par}
		{\sl 
			$^b$~Istituto Nazionale di Fisica Nucleare, Sezione di Padova, I-35131 Padova, Italy
		}\\
		
		{\vskip 1.65cm\par}}
\end{center}

\begin{abstract}

We argue that in general renormalizable field theories the topological angles may develop an additive beta function starting no earlier than 2-loop order. The leading expression is uniquely determined by a single model-independent coefficient. The associated divergent diagrams are identified and a few methods for extracting the beta function in dimensional regularization are discussed. We show that the peculiar nature of the topological angles implies non-trivial constraints on the anomalous dimension of the CP-violating operators and discuss how a non-vanishing beta function affects the Weyl consistency conditions. Some phenomenological considerations are presented.

\end{abstract}

\blfootnote{
	\hypertarget{AV}{\href{mailto:alessandro.valenti@pd.infn.it}{\color{black}{alessandro.valenti@pd.infn.it}}}
}
\blfootnote{
	\hypertarget{LV}{\href{mailto:luca.vecchi@pd.infn.it}{\color{black}{luca.vecchi@pd.infn.it}}}
}

\newpage

{
	\hypersetup{linkcolor=black}
	\tableofcontents
}

\section{Motivations}

Relativistic four-dimensional quantum field theories feature several renormalizable operators: Yukawa, gauge, scalar self-interactions and the topological terms. The latter stand out of this list because they are total derivatives. At the classical level they have no impact since they do not alter the equations of motion; at the quantum mechanical level they do not introduce Feynman vertices. Nevertheless, the topological angles $\theta$ can affect physical observables if non-perturbative effects are taken into account \cite{Belavin:1975fg,tHooft:1976rip,Callan:1976je,Jackiw:1976pf} or if topological defects are present \cite{Witten:1979ey}. In a semi-classical analysis one studies the quantum fluctuations around a certain background field configuration, and $\theta$ shows up in the path integral via a factor $e^{i\theta\nu}$ whenever the background gauge fields are associated to a non-trivial topological charge $\nu$.

Because $\theta$ can appear in observables, it is only natural to wonder about its renormalization group evolution. Since $\theta$ does not parametrize vertices for the quantum fluctuations, it is clear that the perturbative correlators and beta functions cannot depend on it. At most, perturbatively $\theta$ represents a {\emph{counterterm}} necessary to reabsorb (finite and divergent) corrections to CP-violating diagrams with external background gauge fields. In other words, in perturbation theory $\theta$ can only get additively renormalized because of contributions induced by the other couplings.

That $\theta$ renormalizes is out of the question. We can distinguish between {\emph{finite}} and {\emph{infinite}} renormalization effects. Finite (threshold) corrections to $\theta$ are very common. In generic theories they occur at tree-level, when crossing a CP-violating fermion mass threshold, and even more often at loop level. In the case of the QCD $\theta$ angle, for example, ref. \cite{Khriplovich:1985jr} found that the leading order finite correction is a 3-loop effect when matching to the effective field theory below the $W^\pm$ mass. Infinite corrections are more rare. A necessary condition for these contributions to actually occur in a mass independent renormalization scheme is that the quantum field theory under consideration possesses polynomial CP-odd flavor-invariant combinations of the couplings (other than $\theta$) that can contribute to 
\ba
\beta_\theta=\mu\frac{d\theta}{d\mu}.
\ea
In the Standard Model the first such combination appears at a very high power in the Yukawa couplings \cite{Ellis:1978hq,Khriplovich:1993pf} indicating that, if present, the UV divergence that renormalizes $\theta$ should occur at a prohibitive perturbative order. In view of this one may ask if such divergence exists at all, and whether this high perturbative order is a general feature of renormalizable quantum field theories. To better assess these questions it would be useful to find ``toy" field theories in which $\theta$ develops a beta function at a sufficiently low order as to allow an explicit computation, which may provide an indirect confirmation of our expectations in the Standard Model. It is not difficult to find non-renormalizable field theories that induce divergent corrections to $\theta$. For example, adding $cg^2|H|^2G\widetilde G/\Lambda^2$ to the Standard Model Lagrangian gives the minimal subtraction result $\beta_\theta=-4cm_H^2/\Lambda^2$, see \cite{Jenkins:2013zja}. However, the latter effect is power-law suppressed and is therefore practically irrelevant in the presence of a large gap between the IR scales and the UV cutoff. In addition, we will see there are a few non-trivial challenges that a calculation of divergent contributions to $\theta$ faces within renormalizable theories, including the fact that the regularization scheme adopted must be able to consistently deal with the so-called $\gamma_5$ problem. In the non-renormalizable theory we mentioned above those challenges are not encountered because the operator $G\widetilde G$ was already present on the outset. The questions we are interested in therefore better be addressed within {\emph{renormalizable}} field theories. Yet, surprisingly enough, to the best of our knowledge nobody has ever found divergent corrections to $\theta$ in that context.

While no concrete renormalizable example where $\theta$ gets infinitely renormalized is on the market, there exists well-known instances in which one can rigorously prove that $\theta$ {\emph{is not}} infinitely renormalized at any order. An example is pure Yang-Mills, where this property follows trivially because that theory does not satisfy the necessary condition stated above: there are no flavor-invariant CP-odd phases other than $\theta$, and hence there is nothing that can contribute perturbatively to $\beta_\theta$. The same holds for QCD with massive fermions, since once the phases in the quark mass matrix are removed via anomalous chiral rotations of the fermions, CP-violation is entirely encoded in $\theta$. Another popular instance is provided by supersymmetric gauge theories, where the exact one-loop running of the holomorphic gauge coupling reveals that the theta angle does not run. The reason here is again the same: there is no available CP-odd (holomorphic) combination of the other marginal couplings.

The main goals of this paper are to present concrete examples of four-dimensional renormalizable field theories that can induce infinite corrections to $\theta$, as well as to identify the leading order structure of the beta function $\beta_\theta$ in any mass-independent renormalization scheme (see Section \ref{sec:renQFT}); to discuss the subtleties encountered in a perturbative treatment of $\theta$ as well as to show how to concretely approach the calculation of $\beta_\theta$ in dimensional regularization (see Sections \ref{thetaDimReg}); and finally to analyze the relevance of $\beta_\theta$ (Section \ref{sec:implications}).~\footnote{It is worth dissipating a possible source of confusion right away. It is well-known that physical observables must depend on a flavor-invariant combination $\bar\theta$ of the topological angle and the other couplings of the theory (e.g. the quark masses in QCD or the Yukawa couplings in the Standard Model). In a generic field basis the beta function of $\bar\theta$ receives contributions from corrections to both $\theta$, which represent the main subject of this paper, and the other couplings as well (these latter corrections are those estimated in \cite{Ellis:1978hq}, strictly speaking). Throughout the paper we will be mostly concerned with the basis-dependent parameter $\theta$, the coefficient of the topological term. The physics of the QCD parameter $\bar\theta$ will be discussed in Section \ref{sec:UVstrongCP}.
}

Studying the RG evolution of $\theta$ in a general renormalizable theory is not only of academic interest. There are technical as well as potentially phenomenological reasons for doing it. At a genuinely phenomenological level, studying the beta function of $\theta$ in general theories might potentially shed light on the absence of CP-violation in QCD, or more precisely on the necessary properties that its UV completion must satisfy to account for experimental observations, which in fact was the original motivation of our precursors \cite{Wilczek:1977pj,Ellis:1978hq}. Yet another reason is that we do not know what theory will eventually be found to complete the SM at shorter distances. It is perhaps not completely unconceivable that in such a UV-completion topological angles occur in some particular observable, say because of the presence of magnetic monopoles or because small-instanton effects cannot be ignored. In these situations a renormalization group evolution of $\theta$ may become phenomenologically relevant. At a more technical level, $\beta_\theta$ is an obvious target for explicit calculations. The beta functions of the ``ordinary" couplings of general renormalizable field theories have been explicitly calculated up to 2-loop order (see \cite{Machacek:1983tz,Machacek:1983fi,Machacek:1984zw} and more recent updates). Yet, strictly speaking, this program cannot be viewed as fully complete until $\beta_\theta$ is also computed. On a completely different note, the peculiar nature of $\theta$ encodes important information about the renormalization properties of the theory. For example we will see that the independence of perturbative correlators on $\theta$ translates into a constraint on the anomalous dimension of the CP-odd operators of the theory. These and other observations may turn out to be useful in concrete calculations.

\section{$\theta$ in perturbation theory}
\label{thetaDimReg}

In a semi-classical approach to quantum field theory, the bare fields $\phi_0$ are split into a classical finite-action background $\phi_{0c}$ overlapping with the vacuum, plus a quantum fluctuation $\delta\phi_0$ that vanishes sufficiently fast at the boundary. The path integral is defined to include a sum over all inequivalent background configurations (e.g. collective coordinates) along with the functional integration over the fluctuations around each background. Singling out the topological term from the total gauge-fixed action,~\footnote{Throughout the paper we assume the gauge-fixing preserves the background gauge invariance.} which we schematically write as $S_{\rm tot}=S+({g_0^2\theta_0})/({32\pi^2})\int G_0\widetilde G_0$, this recipe produces the following generating functional
\ba\label{ZJ}
{\cal Z}[J_0]
&=&\sum_{\phi_{0c}}~e^{i\frac{g_0^2\theta_0}{32\pi^2}\int G_{0c}\widetilde G_{0c}+i\int J_0\phi_{0c}}\int{\cal D}\delta\phi_0~e^{iS[\phi_{0c}+\delta\phi_0]+i\int J_0\delta\phi_0}\\\no
&=&\sum_{\phi_{0c}}~e^{i\frac{g_0^2\theta_0}{32\pi^2}\int G_{0c}\widetilde G_{0c}}~\widehat{\cal Z}[J_0,\phi_{0c}].
\ea
The topological term is special because it is a total derivative. By construction the quantum fluctuations vanish at the boundary, so that any fluctuation-dependent contribution to such a term vanishes and $\int G_{0}\widetilde G_{0}=\int G_{0c}\widetilde G_{0c}$ reduces to an integral of the sole external background fields, which we can take outside the functional integral in \eqref{ZJ}: the angles $\theta$ do not appear in any interaction of the quantum fluctuations but can act as counterterms in computations with external background fields. 

So far our discussion has been rather general. Yet, an actual evaluation of \eqref{ZJ} is necessarily regularization-scheme dependent. In the following we will specialize on dimensional regularization (Dim-Reg), in which space-time is continued to $d$ dimensions with coordinates $x^\mu=\left\{x^{\bar\mu},x^{\hat\mu}\right\}$, where $\bar\mu,\bar\nu,\cdots=0,1,2,3$ and $\hat\mu,\hat\nu,\cdots$ denote the $(d-4)$-dimensional indices.

In Dim-Reg the very definition of topological term forces us to define the Levi-Civita tensor and deal with the famous $\gamma_5$ problem. So far the only known consistent prescription is the 't Hooft-Veltman-Breitenlohner-Maison scheme~\cite{tHooft:1972tcz,Breitenlohner:1977hr}, where the Levi-Civita tensor is a formal object $\epsilon^{\bar\mu\bar\nu\bar\alpha\bar\beta}$ that carries only 4-dimensional indices. In other words, the $(d-4)$-dimensional indices $\hat\mu,\hat\nu,\cdots$ of an arbitrary vector do not contribute when contracted with this tensor. An important implication is that 
\ba\label{dK}
G_0\widetilde G_0\equiv\frac{1}{2}G_0^{\bar\mu\bar\nu}G_0^{\bar\alpha\bar\beta}\epsilon_{\bar\mu\bar\nu\bar\alpha\bar\beta}=\partial_{\bar\mu} K_0^{\bar\mu}
\ea
is $4$-dimensional divergence of a ($x^\mu$-dependent) vector. Hence the regularized quantity $\int d^dx~G_{0c}\widetilde G_{0c}$ contains a non-trivial residual $(d-4)$-dimensional integral and is not a topological term in $d$ dimensions.

The $d$-dimensional continuation of \eqref{ZJ} formally represents a set of regularized Green's functions. Such a path integral violates two of the familiar properties of the topological angle, namely its periodicity in $2\pi$ and its role as compensator (spurion) of the abelian axial symmetry. \footnote{In our discussion we implicitly assume the theory has integer topological index $\nu$. The extension of our arguments to theories in which $\nu$ is rational is straightforward.} The technical reason for the first loss boils down to the fact that, as a consequence of \eqref{dK}, the bare angle $\theta_0$ is not the coefficient of a topological operator in the regularized theory. The second loss occurs because anomalies are $d$-dependent; as a result, in $d$-dimensions a shift of the coefficient of $G\widetilde G$ does not fully compensate an axial rotation. An intuitive way of arriving to the same conclusions is provided by dimensional analysis: the engineering dimension of the bare coupling in Dim-Reg is $[\theta_0]=d-4$, and it is therefore impossible for $\theta_0$ to be periodic in $2\pi$ or even to shift via the dimensionless parameter of the axial transformation while retaining its $\mu$-independence in $d$-dimensions.

To recover the topological nature of the theta angle, as well as its role as a compensator for abelian axial transformations, one has to derive the renormalized 4-dimensional version of the path integral. In general this procedure requires a renormalization of theta as well. Renormalization renders $\theta$ a genuinely (4-dimensional) topological term and the background-dependence in the 4-dimensional limit of the path integral \eqref{ZJ} reduces to a dependence on the topological index $\nu$. The renormalized coupling $\theta$ is periodic in $2\pi$ and transforms via a shift under abelian axial rotations. This ensures in particular that physical amplitudes are invariant under unitary field re-definitions.

\subsection{$\beta_\theta$ from (extra)-ordinary diagrams}

For completeness we recall the standard prescription to extract the beta function within the Minimal Subtraction scheme. The relation between the bare couplings $\theta_0$, $\xi_{0i}$ (the latter symbol denotes all couplings except $\theta_0$) and the renormalized couplings $\theta$ and $\xi_i$ read (in $d=4-\epsilon$ dimensions)
\ba\label{divZtheta}
\theta_0=\mu^{-\epsilon}[\theta +Z_\theta],
\ea
and $\xi_{0i}=\mu^{\rho_i\epsilon}[\xi_i+Z_{\xi_i}]$. By definition $Z_{\theta}=\sum_{n=1}^\infty\epsilon^{-n} Z_{\theta,n}(\xi_i)$ contains no finite term, and similarly for $Z_{\xi_i}$. The $4$-dimensional beta functions $\beta_\xi\equiv\lim_{d\to4}\mu{d\xi}/{d\mu}$ read $\beta_\theta=\rho_i\xi_i{\partial Z_{\theta,1}}/{\partial\xi_i}+ Z_{\theta,1}$ and similarly $\beta_{\xi_i}=\rho_j\xi_j{\partial Z_{i,1}}/{\partial\xi_j}-\rho_i Z_{i,1}$. In the above $\partial Z_{\theta}/\partial\theta=\partial Z_{\xi_i}/\partial\theta=0$ because $\theta$ does not appear in Feynman diagrams. As customary for ordinary couplings, also the beta function of $\theta$ is controlled by the simple pole $Z_{\theta,1}/\epsilon$. Diagrammatically, the latter counterterm is defined to subtract the divergent contributions to the connected, CP-odd {\emph{vertices}} with external background gauge fields in $\widehat{\cal Z}[J_0,\phi_{0c}]$, see \eqref{ZJ}:
\ba\label{ciocheciint}
\widehat{\cal Z}[J_0,\phi_{0c}]\supset-\frac{Z_{\theta,1}}{\epsilon}i\frac{g_0^2}{32\pi^2}\mu^{-\epsilon}\int d^dx\, G_{c0}\widetilde G_{c0}.
\ea
The divergent corrections to the external legs are removed via a renormalization of the background field, $A^\mu_{0c}=(g/g_0)A^\mu_c$, so that $g_0^2G_{0c}\widetilde G_{0c}=g^2G_{c}\widetilde G_{c}$. The divergence remaining in \eqref{ciocheciint}, if any, must be subtracted by $Z_{\theta,1}$.

A word regarding the actual computation of $Z_{\theta,1}$ is now needed. After all, we have argued that the topological term does not represent an ordinary vertex, so how can we now claim that combining the ordinary vertices of the theory one can find divergent corrections to it? The key point is that \eqref{ciocheciint}, in order to be non-vanishing, must be a functional of classical backgrounds with {\emph{non-trivial}} asymptotic behavior. This implies the calculation of $Z_{\theta,1}$ must be dealt with care. In particular, integration by parts cannot be performed lightly, as opposed to what is customary done when dealing with external sources for asymptotic states.~\footnote{If integration by parts is performed neglecting boundary terms, the structure \eqref{ciocheciint}, say the part with two external legs for definiteness, would be written in the familiar form 
\ba\label{intbypar}
\widehat{\cal Z}[J_0,\phi_{0c}]\supset\int d^dx A_{0c}^\mu(x)\int d^dy A_{0c}^\nu(y)~G_{\mu\nu}(x-y), 
\ea
with $G_{\mu\nu}(x-y)$ an ordinary Feynman diagram. This cannot contribute to the beta function of $\theta$, since it is impossible for $G_{\mu\nu}(x-y)$ to contain two derivatives contracted with a Levi-Civita tensor, as required to reproduce the topological term in \eqref{ciocheciint}. Any CP-violating structure of the type \eqref{intbypar} identically vanishes and cannot describe \eqref{ciocheciint}.} One way to proceed is to extract the divergent piece in \eqref{ciocheciint} directly in position space while allowing an arbitrary background $A^\mu_c$ \cite{Novikov:1983gd}. This is the approach followed in \cite{Khriplovich:1985jr} and \cite{Pospelov:1994uf} to compute the induced $\theta$ term in the Standard Model below the weak scale.

An alternative way to calculate the divergent diagrams contributing to $Z_{\theta,1}$ may be via the trick proposed in \cite{Georgi:1980cn}. The idea is to promote the CP-odd couplings to non-propagating fields, i.e. ``axions", in the intermediate steps and then send them to constant values at the very end of the computation. In this way $\int d^dx~\theta\,G\widetilde G$ would describe an ordinary vertex with $\theta(x)$ and a number of gluons, which is non-vanishing as long as the external $\theta(x)$ carries a momentum. It is now plausible that ordinary Feynman diagrams with external gluons and non-dynamical axions contain divergent contributions that need to be subtracted by $\int d^dx~\theta\,G\widetilde G$, very much like ordinary diagrams were shown in \cite{Georgi:1980cn} to be capable of generating finite threshold corrections to $\theta$. Amusingly, though, now that the topological term behaves as an ordinary vertex one may naively expect $\theta$ to be able to show up in matrix elements as well as in the beta functions. It turns out however that this cannot happen. One way to see it is that the renormalized S-matrix amplitudes must be periodic functions of $\theta$. Yet, the new vertex described by the topological term is measured by the strength $\theta g^2/32\pi^2$. Hence the only way it can contribute to renormalized amplitudes and beta functions is via powers of $\theta g^2/32\pi^2$. However there is no way that a perturbative function of $\theta g^2/32\pi^2$ and the other couplings be invariant under $\theta\to\theta+2\pi$ unless the dependence on $\theta g^2/32\pi^2$ is actually absent altogether. The situation is completely different when non-perturbative effects are taken into account, of course, since in that case inverse powers of the gauge coupling cannot be excluded a priori and a dependence on $\theta g^2/32\pi^2$ can be turned into a periodic function of the sole $\theta$.

We see that the divergent contributions to $\theta$ can be seen via non-standard perturbative calculations, either by directly evaluating \eqref{ciocheciint} in position space \cite{Novikov:1983gd} or via the method proposed in \cite{Georgi:1980cn}. In Section \ref{sec:anomalDIM} we will present a third alternative approach to calculate directly $\beta_\theta$, based on more conventional perturbative methods. 

Yet, whatever method is adopted one has to pay particular attention to how chirality is implemented. The divergent contribution we are interested in is proportional to the Levi-Civita tensor and must arise from a fermion trace involving the $\gamma_5$ matrix. A consistent treatment of the latter has to be implemented, and at present in Dim-Reg the unique option seems to be given by the 't Hooft-Veltman-Breitenlohner-Maison scheme. Unfortunately, using this scheme might complicate the calculation a bit, due to the non-standard anti-commutation properties that $\gamma_5$ satisfies and the necessity of introducing ad-hoc symmetry-restoring counterterms in order to preserve non-anomalous (global and gauge) chiral symmetries. Fortunately the results of this paper are completely general and hold independently of the approach followed.

\subsection{Extracting $\beta_\theta$ from operator mixing}
\label{sec:anomalDIM}

The insertion of composite operators in correlator functions can be systematically described by introducing in the bare Lagrangian appropriate spacetime-dependent sources for them (and adding the appropriate counterterms). After having integrated out the dynamical fields one is left with a generating functional for the time-ordered, connected correlation functions of the renormalized operators. The local renormalization group \cite{Osborn:1989td,Jack:1990eb,Osborn:1991gm,Jack:2013sha} is a very powerful incarnation of this general prescription. In that approach the Lagrangian is expanded in a complete basis of operators ${\cal O}$ and all couplings of the theory become functions of spacetime, including the metric that sources the energy momentum tensor. A local version of the renormalization group equation for the operators ${\cal O}$ can be derived. Once all couplings are sent to their constant ($x^\mu$ independent but $\mu$-dependent) background values, this reads \cite{Baume:2014rla}
\ba\label{localRGCS}
\mu\frac{d}{d\mu}\left(\begin{matrix} \partial_\mu J_{\dot A}^\mu\\ {\cal O}_{\bar I}\end{matrix}\right)=-\left(\begin{matrix} \gamma_{\dot A\dot B} & 0\\\gamma_{\bar I\dot B} & \frac{\partial}{\partial \xi_{\bar I}}\beta_{{\bar J}}\end{matrix}\right)\left(\begin{matrix} \partial_\mu J_{\dot B}^\mu\\ {\cal O}_{\bar J}\end{matrix}\right).
\ea
Here $J_{\dot A}^\mu$ are the (conserved or anomalous \cite{Keren-Zur:2014sva}) currents of the global symmetries of the theory, ${\cal O}_{\bar I}$ denote the marginal interactions, $\xi_{\bar I}$ the associated couplings, and $\beta_{{\bar I}}=\mu d\xi_{\bar I}/d\mu$. We considered a theory without mass operators for simplicity, but the introduction of mass terms is straightforward and does not affect our discussion.~\footnote{
A complete basis of ${\cal O}_{\bar I}$'s includes two sets of marginal operators: ${\cal O}_{\bar I}=\left\{E_{\bar I''},O_{\bar I'}\right\}$. The interactions $O_{\bar I'}$ are those that define the bare action. These are multiplied by sources $\xi_{\bar I'}$ whose background values represent the ordinary couplings of the theory. The $E_{\bar I''}$'s denote instead redundant marginal interactions. They include evanescent operators as well as operators that vanish via the equations of motion. They are not associated to any actual coupling of the theory. However they must be included in the functional integral multiplied by spacetime-dependent sources $\xi_{\bar I''}$ in order for ${\cal O}_{\bar I}$ to form a closed set of composite operators under renormalization. The background value of such sources vanishes. Because the $\xi_{\bar I''}$'s are not actual couplings, their (background-value) beta functions are proportional to the $\xi_{\bar I''}$'s themselves times functions of the true couplings $\xi_{\bar I'}$, i.e. the operators $E_{\bar I''}$ are multiplicatively renormalized. With this observation we see that \eqref{localRGCS} describes a familiar pattern of RG-mixing: the anomalous dimension matrix for $\left\{E_{\bar I''},O_{\bar I'}\right\}$ has a lower triangular form in which the redundant operators renormalize among each other whereas the $O_{\bar I'}$ renormalize via a mixture of themselves, $\partial_\mu J_{\dot A}^\mu$, and $E_{\bar I''}$.}

Let us begin by discussing the consequence of \eqref{localRGCS} on the CP-odd sector of QCD. The marginal CP-odd operators are the divergence of the singlet axial current, $\partial_\mu J_5^\mu$, and the topological term $G\widetilde G$ (modulo operators that vanish via the equations of motion). By CP-invariance the anomalous dimension matrix contains a block diagonal 2 by 2 subgroup involving these two operators only (the operators that vanish on-shell do not affect the following discussion). From \eqref{localRGCS} one confirms that this takes the form discussed in \cite{Espriu:1982bw}. The vanishing $\gamma_{\dot A\bar J}=0$ entry is understood as a result of the fact that in Dim-Reg the current $J_5^\mu$ renormalizes multiplicatively, being the unique gauge-invariant axial current of the theory, and the same is true for its derivative. Eq. \eqref{localRGCS} reveals that the $G\widetilde G$-$G\widetilde G$ element of the anomalous dimension, often denoted by $\gamma_{G\widetilde G}$ in the literature, is given by $\gamma_{G\widetilde G}={\partial}\beta_{{G\widetilde G}}/{\partial \xi_{G\widetilde G}}$, with $\xi_{G\widetilde G}$ being the renormalized coupling
\ba
\xi_{G\widetilde G}\equiv\frac{\theta g^2}{32\pi^2}.
\ea
Crucially, $\theta$ cannot appear in any perturbative calculation, so all our expressions are understood as being evaluated at $\xi_{G\widetilde G}=0=\theta$. As expected, in the external source formalism the latter can only occur with derivatives \cite{Shore:1990wp}. Because $\mu d({\theta g^2})/d\mu=\beta_\theta g^2+\theta\beta_{g^2}$ we find
\ba\label{GenLarin}
\gamma_{G\widetilde G}=\left.\frac{\partial}{\partial \xi_{G\widetilde G}}\beta_{{G\widetilde G}}\right|_{\xi_{G\widetilde G}=0}=\left.\frac{\partial}{\partial \xi_{G\widetilde G}}\left[\beta_{\theta}\frac{g^2}{32\pi^2}+\xi_{G\widetilde G}\frac{\beta_{g^2}}{g^2}\right]\right|_{\xi_{G\widetilde G}=0}=\frac{\beta_{g^2}}{g^2}.
\ea
In the last equality it is crucial that $\beta_\theta$ nor $\beta_{g^2}$ depend on $\theta$, or equivalently $\xi_{G\widetilde G}$. The resulting relation between $\gamma_{G\widetilde G}$ and the beta function of the gauge coupling is consistently observed in explicit calculations \cite{Larin:1993tq,Ahmed:2021spj}~\footnote{We use a different convention than these authors. For us the scaling dimension of an operator is $d_{\rm cl}+\gamma$, with $d_{\rm cl}$ the engineering dimension.} and proven in \cite{Breitenlohner:1983pi,Luscher:2021bog}. Note that $\beta_\theta$ disappears from this relation and \eqref{GenLarin} would still hold even if it was non-trivial. Yet, we know that $\beta_\theta=0$ in QCD because of the argument given in the introduction: by CP-invariance of the theory the only quantity that could appear in $\beta_\theta$ is $\theta$ itself, but this can never happen in perturbation theory.

Suppose now we extend QCD introducing in the Lagrangian new CP-violating dimension-4 operators $O_{\bar I}$ with $\bar I\neq G\widetilde G$, e.g. Yukawa couplings. A priori these may mix with the topological term as well as the derivative of the axial current. Yet, according to \eqref{localRGCS} the anomalous dimension matrix has a lower triangular form. This is because by dimensional analysis the renormalized axial and Chern-Simons currents cannot contain a component of the bare dimension-4 interactions $O_{\bar I0}$, they can only mix among each other following the pattern described above. On the other hand, nothing forbids the renormalized $O_{\bar I}$ to contain a linear combination of $\partial_\mu J_{5\,0}^\mu, G_0\widetilde G_0$. The lower-triangular structure can also be understood as a consequence of the independence of the anomalous dimensions and beta functions on $\theta$, i.e. $\partial\beta_{\bar I\neq G\widetilde G}/\partial\xi_{G\widetilde G}=0$ \cite{Keren-Zur:2014sva}. 

Interestingly, eq. \eqref{localRGCS} demonstrates that \eqref{GenLarin} remains valid even in the presence of the new interactions $O_{\bar I}$ (${\bar I}\neq G\widetilde G$). More importantly for us, the new CP-violating couplings make it possible for $\theta$ to get additively renormalized, and \eqref{localRGCS} turns out to contain important information on $\beta_\theta$. The component of $-\mu dO_{\bar I}/d\mu$ proportional to $G\widetilde G$ is (for $\bar I\neq G\widetilde G$)
\ba\label{betaAD}
\gamma_{{\bar I}G}=\left.\frac{\partial}{\partial \xi_{\bar I}}\beta_{{G\widetilde G}}\right|_{\xi_{G\widetilde G}=0}=\left.\frac{\partial}{\partial \xi_{\bar I}}\left[\beta_{\theta}\frac{g^2}{32\pi^2}+\xi_{G\widetilde G}\frac{\beta_{g^2}}{g^2}\right]\right|_{\xi_{G\widetilde G}=0}=\frac{g^2}{32\pi^2}\frac{\partial}{\partial \xi_{\bar I}}\beta_{\theta}.
\ea
Eq. \eqref{betaAD} may be interpreted as an indirect procedure for deriving $\beta_\theta$. The latter may indeed be extracted by integrating the off-diagonal element $\gamma_{\bar I G}$ of the anomalous dimension matrix of the CP-violating operators, roughly measured by the amount of $G_0\widetilde G_0$ contained in $O_{\bar I}$, with respect to the couplings $\xi_{\bar I}$ that contribute additively to $\beta_\theta$. Interestingly, this procedure does not require the background field method nor promoting the CP-odd phases to spurion fields. It can be carried out using ordinary perturbation theory because the anomalous dimension matrix is to be calculated assuming a non-vanishing momentum is flowing into the relevant operators, such that even $G\widetilde G$ describes an ordinary vertex. 

A generalization of eq. \eqref{betaAD} to arbitrary renormalizable theories is straightforward.

\section{$\beta_\theta$ in renormalizable QFTs}
\label{sec:renQFT}

In the previous section we have shown how to extract the beta function of the topological angle. In this section we are interested in identifying which renormalizable theories have enough CP-violating phases to allow, at least in principle, the presence of divergent contributions to $\theta$. More precisely, we will identify the explicit form of the leading non-trivial order beta function $\beta_\theta$ as well as the Feynman diagrams responsible for generating it.

The most general ($d=4$) renormalizable gauge theory can be compactly written in terms of Weyl fermions $\psi_i$ and real scalars $\phi_a$ as
\ba\label{LagGen}
{\cal L}
&=&-\frac{1}{4g^2_{AB}}F^A_{\mu\nu}F^{B\mu\nu}+\frac{1}{2}(D_\mu\phi)_a(D_\mu\phi)_a+\psi^\dagger_i i\bar\sigma^\mu ({D}_\mu\psi)_i\\\no
&-&\left(\frac{1}{2}Y_{a \, ij}\psi_i\psi_j\phi_a+{\rm hc}\right)-\frac{\lambda_{abcd}}{4!}\phi_a\phi_b\phi_c\phi_d+\frac{\theta^{AB}}{64\pi^2}\epsilon^{\mu\nu\rho\sigma}F_{\mu\nu}^A F_{\rho\sigma}^B\\\no
&+&({\rm relevant~couplings})+({\rm gauge~fixing})+({\rm ghosts}),
\ea
where $(D_\mu\psi)_i=\partial_\mu\psi_i-i{A}^A_\mu T^A_{ij}\psi_j$ and $(D_\mu\phi)_a=\partial_\mu\phi_a-i{A}^A_\mu S^A_{ab}\phi_b$. The gauge generators $T^A$ are hermitian whereas $S^A$ are purely imaginary hermitian, and hence anti-symmetric. The fermions and scalars are in general in a reducible representation of the gauge group, and the indices $i,j,\cdots$ (ranging from $1$ to some integer $N_\psi$) and $a,b,\cdots$ (ranging from $1$ to $N_\phi$) include both gauge and flavor components. The coupling $\lambda$ is fully symmetric, and $Y$ is symmetric in the fermionic indices. The gauge symmetry is an arbitrary product of abelian factors and simple groups ${\cal G}_{\rm gauge}=\Pi_{\cal G}{\cal G}$. The indices $A,B,\cdots$ run over the adjoint representation of ${\cal G}_{\rm gauge}$ and the (real) gauge coupling $g^2_{AB}$ is to be interpreted as the direct sum of identities in the non-abelian part of that space plus a symmetric part for the abelian factors. Note that our normalization of the gauge fields is non-canonical, so that the gauge boson propagators are proportional to $g_{AB}^2$.

Gauge invariance restricts the form of some of the couplings in \eqref{LagGen}. Most importantly for us, the topological term must be proportional to the identity $\delta^{AB}_{\cal G}$ in any non-abelian components ${\cal G}$ and symmetric, but potentially with off-diagonal entries, in the abelian factors:
\ba\label{thetaGEN}
\theta^{AB}=\sum_{{\cal G}={\rm non-Ab.}}\theta_{\cal G}\delta^{AB}_{\cal G}+\sum_{{\cal G}_1,{\cal G}_2={\rm Ab.}}\theta_{{{\cal G}_1,{\cal G}_2}}^{AB}.
\ea
Another implication of gauge invariance is 
\ba\label{SYasYT}
S^A_{ab}Y_b=Y_aT^A+(T^A)^*Y_a.
\ea
We will make use of this relation extensively throughout the text. There is also a constraint on $\lambda$, though such a relation will not be relevant here. Note that none of these relations depend on the space-time dimension and hence can be assumed to hold in Dim-Reg as well.

\subsection{CP and flavor symmetry}

We want to identify which combinations of the couplings can appear in $\beta_\theta$. In Dim-Reg the renormalization group evolution of $\theta$ is necessarily controlled by the marginal couplings of the theory. Classically relevant interactions are therefore not important in the present discussion and have not been explicitated in \eqref{LagGen}. To proceed we first have to familiarize with the approximate symmetries of our theory.

In the absence of interactions Eq. \eqref{LagGen} enjoys a large global symmetry that includes a flavor symmetry that rotates the matter fields, one that rotates vectors, and CP:
\ba\label{FullG}
{\cal G}_{\rm flav}\times {\cal G}_{\rm glob}\times CP.
\ea
The flavor symmetry ${\cal G}_{\rm flav}\equiv U(N_\psi)\times O(N_\phi)$ rotates all fermions among each other and acts similarly on scalars, the group ${\cal G}_{\rm glob}$ rotates the vectors leaving $g_{AB}^2$ in \eqref{LagGen} invariant and finally CP acts as usual up to unitary rotations. The combined action on the fields reads:
\ba\label{fieldsG}
\psi_i(x)\to U_{ij}\epsilon\psi^\dagger_j({\cal P}x),~~~~\phi_a(x)\to O_{ab}\phi_b({\cal P}x),~~~~A^A_\mu(x)\to R^{AB}{\cal P}_\mu^\nu A^B_\nu({\cal P}x)
\ea
where $U,O,R$ are matrices respectively of $U(N_\psi)$, $O(N_\phi)$, ${\cal G}_{\rm glob}$ and we defined ${\cal P}_\mu^\nu x_\nu=x^\mu$. The symmetry \eqref{FullG} is explicitly broken by the gauge generators, the Yukawa and scalar couplings as well as anomalies. Yet, it can be formally restored by promoting $T^A$, $S^A$, $Y_a$, $\lambda, \theta$ to spurions with transformations designed to exactly compensate \eqref{fieldsG} so that the theory is manifestly invariant under the full group \eqref{FullG}.~\footnote{In Dim-Reg the addition of counterterms is generically necessary to render the theory formally invariant under the axial part of $U(N_\psi)$. We assume this is done order by order in perturbation theory.} Explicitly, the transformations of the spurions $T^A$, $S^A$, $Y_a$, $\lambda$ under ${\cal G}_{\rm flav}\times {\cal G}_{\rm glob}\times CP$ are given by
\ba\label{symmGEN}
T^A_{ij}&\to& -R^{AB}U_{im}U^*_{jn}[T^B]^*_{mn}\\\no
S^A_{ij}&\to& -R^{AB}O_{im}O_{jn}[S^B]^*_{mn}\\\no
Y_{a \, ij}&\to& U^*_{im}U^*_{jn}O_{ab}[Y_{b \, mn}]^*\\\no
\lambda_{abcd}&\to& O_{am}O_{bn}O_{co}O_{dp}\lambda_{mnop}.
\ea
The renormalized coupling $\theta^{AB}$ should also be interpreted as a spurion. It is a complete singlet of $O(N_\phi)$, a 2-index symmetric tensor of ${\cal G}_{\rm glob}$, and is also $U(N_\psi)$-invariant except for its anomalous subgroup. Denoting with $T_r^A$ the generators of each irreducible representation $r$ of the gauge group modulo flavor degeneracies, and by $U_r$ the flavor rotation among the fermions in $r$, the spurious ${\cal G}_{\rm flav}\times {\cal G}_{\rm glob}\times CP$ transformation of $\theta$ explicitly reads
\ba\label{thetaCP}
\theta^{AB}\to-R^{AM}R^{BN}\left\{\theta^{MN}-2\sum_r{\rm Arg}~{\rm Det}[U_r]~{\rm Tr}[T_r^MT_r^N]\right\}.
\ea
Here the overall minus is necessary to compensate the P-odd nature of $\epsilon^{\mu\nu\rho\sigma}F_{\mu\nu}^A F_{\rho\sigma}^B$.

The beta function $\mu d\theta^{AB}/d\mu$ is not affected by the anomalous shift of the topological term and therefore is a complete singlet of the flavor symmetry ${\cal G}_{\rm flav}$ and CP-odd in the sense that 
\ba\label{betaCP}
\mu \frac{d}{d\mu}\theta^{AB}\to-R^{AM}R^{BN}\mu \frac{d}{d\mu}\theta^{MN}. 
\ea
The beta function is a linear combination of functions $I^{AB}$ of the spurions \eqref{symmGEN} multiplied by real numerical coefficients, since its calculation involves no branch cuts. The quantities $I^{AB}$ are obtained by contracting the indices of the spurions with the invariant tensors $\delta_{ij},\delta_{ab},g^2_{AB}$ that define the kinetic terms. In the language of Feynman diagram, this is just a consequence of the fact that any diagram contributing to the beta function is obtained by contracting vertices with propagators. In order to appear in $\beta_\theta$ the tensors should transform precisely as in \eqref{betaCP}. Yet, from \eqref{symmGEN} follows that under the full symmetry \eqref{FullG} any 2-index tensor function of \eqref{symmGEN} transforms as 
\ba\label{ICP}
I^{AB}(T,S,Y,\lambda)\to R^{AM}R^{BN}I^{MN}(T^*,S^*,Y^*,\lambda^*)
\ea
The minus signs of  \eqref{symmGEN} cancel out because ${\cal G}_{\rm glob}$-covariance forces $I^{AB}$ to be built out of an even number of (scalar plus fermion) gauge generators. Now, to reproduce the transformation of $\mu d\theta^{AB}/d\mu$ the tensors must satisfy $I^{AB}(T^*,S^*,Y^*,\lambda^*)=-I^{AB}(T,S,Y,\lambda)$, and of course be real. We thus conclude that it is the imaginary part of the invariant that has the correct CP-odd property, namely~\footnote{The relations $I^{AB}(T^*,S^*,Y^*,\lambda^*)=[I^{AB}(T,S,Y,\lambda)]^*=-I^{AB}(T,S,Y,\lambda)$, the first equality following from the fact that all coefficients are real and the second from the requirement that \eqref{ICP} reproduces \eqref{betaCP}, are equivalent to ${\rm Re}[I^{AB}(T^*,S^*,Y^*,\lambda^*)]={\rm Re}[I^{AB}(T,S,Y,\lambda)]=0$ and ${\rm Im}[I^{AB}(T^*,S^*,Y^*,\lambda^*)]=-{\rm Im}[I^{AB}(T,S,Y,\lambda)]$.}
\ba
\mu\frac{d}{d\mu}\theta^{AB}=\sum_\alpha c_\alpha~ {\rm Im}\left[I_{\alpha}^{AB}\right],
\ea
where $c_\alpha$ are real numbers and $\alpha$ is some label. In other words, the CP-odd invariants that define $\beta_\theta$ are the {\emph{imaginary}} parts of the ${\cal G}_{\rm flav}$-singlet, 2-index tensors of ${\cal G}_{\rm glob}$. A theory that does not possess any such quantity cannot renormalize $\theta$. This is what happens in Yang-Mills theories as well as QCD. In the next subsection we will show the explicit form of the leading order contribution to $\beta_\theta$ in arbitrary renormalizable theories of the form \eqref{LagGen}.

\subsection{The 3-loop diagrams}

Following the method described above we identified all structures that can potentially contribute to $\beta_\theta$ at the first few perturbative orders. Our analysis is summarized in Appendix \ref{AppIAB}. We find that there are no 1-loop-sized 2-index tensors that are CP-odd, and therefore that the 1-loop beta function must vanish. The first non-trivial contribution to $\beta_\theta$ potentially arises at 2-loops and is controlled by a unique structure
\ba\label{betaTheta}
\mu\frac{d}{d\mu}\theta^{AB}=c\frac{\hbar^2}{(16\pi^2)^2}{\rm Im}\left[I^{AB}_{(2)}\right]+{\cal O}(\hbar^3),
\ea
where $c$ is an ordinary number expected to be of order unity and
\begin{align}
	\begin{aligned}
		I_{(2)}^{AB}&=\tr \left\{ (Y_a^*[T^A]^*Y_bY_a^*Y_b-Y^*_aY_bY^*_a [T^{A}]^*Y_b) T^B\right\} \\
		&= \tr \left\{ (Y^*_aY_c Y_b^*Y_c-Y^*_cY_a Y_c^*Y_b)T^B  \right\} S^A_{ab}\\
		&=\frac{1}{2}  \tr \left\{ Y^*_aY_c Y_b^*Y_c  -  Y^*_cY_a Y_c^*Y_b\right\} \, (S^A S^B )_{ab}
	\end{aligned}
\label{CPoddInvariantFIN}
\end{align}
is the only invariant with non-vanishing CP-odd component (i.e. an imaginary part) at ${\cal O}(\hbar^2)$. Note that symmetry under the exchange $A\leftrightarrow B$ in the first and third lines of \eqref{CPoddInvariantFIN} is a consequence of cyclicity of the trace as well as symmetry of the Yukawa under the exchange of the fermionic indices and hermiticity of the gauge generators, whereas in the second line of \eqref{CPoddInvariantFIN} also \eqref{SYasYT} is needed.

The invariant $I^{(2)}_ {AB}$ has been written in three different forms employing the identity \eqref{SYasYT}. These expressions provide complementary information about the properties that a theory must possess in order to renormalize $\theta^{AB}$ at this order. For instance, from the second and third lines in \eqref{CPoddInvariantFIN} it is evident that there would be no 2-loop beta function if the scalars were gauge-singlets, i.e. if $S^A=0$. To show this using the expression in the first line is less immediate: one has to use \eqref{SYasYT} with $S^A=0$ to prove that $Y_a^*Y_bT^A=T^AY_a^*Y_b$, from which consistently follows that the first line of \eqref{CPoddInvariantFIN} vanishes. Actually, a more careful inspection reveals that {\emph{the scalar fields have to belong to at least two different representations of $\mathcal{G}$}. To prove this we distinguish between non-abelian and abelian gauge groups. In the case the indices $A,B$ are associated to a non-abelian gauge group, from \eqref{thetaGEN} we know that the relevant part of the beta function is the one proportional to $\delta^{\cal G}_{AB}$. If all the scalars belonged to the same representation, then contracting the latter with the expression in the third line of \eqref{CPoddInvariantFIN} would give a Casimir times the identity in the scalar index space. This implies that the invariant would vanish as $\tr \left\{ Y^*_aY_c Y_a^*Y_c  -  Y^*_cY_a Y_c^*Y_a\right\}=0$. In the case $A,B$ refer to abelian gauge groups the generators can all be taken diagonal, i.e. $S^A_{ab}=q^A_a\delta_{ab}$ and so $(S^A S^B )_{ab}=q^A_aq^B_a\delta_{ab}$. As before, we see that when the charges $q_a^{A},q^B_a$ do not depend on the index $a$ then the combination of scalar generators is again proportional to the identity $\delta_{ab}$ and the third line of \eqref{CPoddInvariantFIN} is identically zero.

The three equivalent forms of \eqref{CPoddInvariantFIN} also help us identify the diagrams that contribute to the 2-loop beta function, e.g. if we adopted the background field method. These are illustrated in Fig. \ref{fig:diagrams}, with the $\otimes$ indicating the insertion of the external background gauge field. The topology in Fig. \ref{fig:diagrams1} is responsible for generating the invariant in the first line of \eqref{CPoddInvariantFIN}, the one in Fig. \ref{fig:diagrams2} is associated to the second line of \eqref{CPoddInvariantFIN} and finally the topology of Fig. \ref{fig:diagrams3} to the third form of the CP-odd invariant. The overall correction to $\beta_\theta$ must include a sum over all three topologies. We emphasize that these are effectively 3-loop diagrams, and yet they contribute to the 2-loop order beta function because $\theta$ appears in the action multiplied, in the canonically normalized field basis, by a loop factor $g^2\hbar/32\pi^2$.

\begin{figure}[!t]
	\begin{center}
		\begin{subfigure}[t]{0.22\textwidth}
		\resizebox{4cm}{!}{
			\begin{tikzpicture}
				\draw (4,4) circle (4cm);
				\draw[dash pattern=on 8pt off 8pt] (0,4) -- (3,4);
				\draw[dash pattern=on 8pt off 8pt] (5,4) arc (0:180:1cm);
				\draw[dash pattern=on 8pt off 8pt] (5,4) -- (8,4);
				\draw[dash pattern=on 8pt off 8pt] (4,0) -- (4,8);
				\draw[black, fill=white] (1.35425cm,7) circle (0.4cm);
				\draw (1.35425cm,7) node[cross=0.3cm,rotate=90] {};
				\draw[black, fill=white] (6.64575cm,7) circle (0.4cm);
				\draw (6.64575cm,7) node[cross=0.3cm,rotate=90] {};
			\end{tikzpicture}
		}
		\caption{\label{fig:diagrams1}\hspace*{-0.75cm}}
	\end{subfigure}
	\hspace{2cm}
	\begin{subfigure}[t]{0.22\textwidth}
		\resizebox{4cm}{!}{
			\begin{tikzpicture}
				\draw (4,4) circle (4cm);
				\draw[dash pattern=on 8pt off 8pt] (0,4) -- (3,4);
				\draw[dash pattern=on 8pt off 8pt] (5,4) arc (0:180:1cm);
				\draw[dash pattern=on 8pt off 8pt] (5,4) -- (8,4);
				\draw[dash pattern=on 8pt off 8pt] (4,0) -- (4,8);
				\draw[black, fill=white] (1.35425cm,7) circle (0.4cm);
				\draw (1.35425cm,7) node[cross=0.3cm,rotate=90] {};
				\draw[black, fill=white] (4,6) circle (0.4cm);
				\draw (4,6) node[cross=0.3cm,rotate=90] {};
			\end{tikzpicture}
		}
		\caption{\label{fig:diagrams2}\hspace*{-0.75cm}}
	\end{subfigure}
	\hspace{2cm}
	\begin{subfigure}[t]{0.22\textwidth}
		\resizebox{4cm}{!}{
			\begin{tikzpicture}
				\draw (4,4) circle (4cm);
				\draw[dash pattern=on 8pt off 8pt] (0,4) -- (3,4);
				\draw[dash pattern=on 8pt off 8pt] (5,4) arc (0:180:1cm);
				\draw[dash pattern=on 8pt off 8pt] (5,4) -- (8,4);
				\draw[dash pattern=on 8pt off 8pt] (4,0) -- (4,8);
				\draw[black, fill=white] (4,2) circle (0.4cm);
				\draw (4,2) node[cross=0.3cm,rotate=90] {};
				\draw[black, fill=white] (4,6) circle (0.4cm);
				\draw (4,6) node[cross=0.3cm,rotate=90] {};
			\end{tikzpicture}
		}
		\caption{\label{fig:diagrams3}\hspace*{-0.75cm}}
	\end{subfigure}
	\hfill\mbox{}
	\end{center}
	\caption{\label{fig:diagrams}Topologies of the diagrams generating the CP-odd invariant \eqref{CPoddInvariantFIN}. Crossed circles represent insertions of the external gauge field. Topology (a) refers to the form in \eqref{CPoddInvariantFIN} written in terms of two fermionic generators, topology (b) to the one in terms of one fermionic and one scalar generator, and topology (c) to the one in terms of two scalar generators. It is intended that each diagram within a given topology should be properly symmetrized.}
\end{figure}
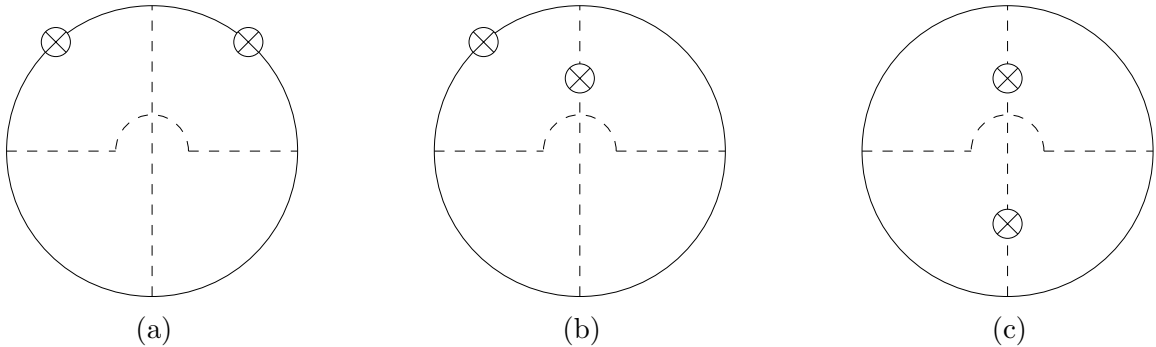

The coefficient $c$ in \eqref{betaTheta} is model-independent and may be derived calculating the above diagrams in any model with a non-vanishing $I_{(2)}^{AB}$, as for example in the toy scenario discussed below. We will not be able to compute it here because far beyond our current technical abilities. Yet, we think there is circumstantial evidence that an explicit computation would find $c\neq0$. Indeed, ref. \cite{Pickering:2001aq} presents a calculation of the beta function of the gauge coupling at 3-loop order using the background field method. The class of diagrams considered there includes those of Fig. \ref{fig:diagrams}.~\footnote{In practice the calculation is very different, though, because here we are interested in the CP-odd contributions proportional to the Levi-Civita tensor (see the discussion in Section \ref{thetaDimReg}), whereas in \cite{Pickering:2001aq} the authors were allowed to take $\epsilon_{\bar\mu\bar\nu\bar\alpha\bar\beta}=0$.} Consistently, those authors find that the beta function of the gauge coupling is controlled by the set of CP-even invariants of Appendix \ref{AppIAB}, including in particular the CP-even component of our $I_{(2)}^{AB}$ (namely its real part). This gives us confidence that the evaluation of the diagrams in Fig. \ref{fig:diagrams} will not cancel against each other. The result $c=0$ would thus be rather surprising to us, and would indicate the presence of an accidental cancellation.

\section{Implications}
\label{sec:implications}

In the introduction we mentioned a few implications of $\beta_\theta$, both technical and phenomenological. Here we elaborate on two of them. First we show the impact of a non-vanishing $\beta_\theta$ on the Weyl consistency conditions. Next we discuss the relevance of $\beta_\theta$ on UV solutions of the strong CP problem.

\subsection{Consistency conditions}
\label{sec:consistencyREL}

Jack and Osborn pointed out that the beta functions in a general P-conserving renormalizable theory must satisfy a constraint (see e.g. \cite{Jack:2013sha})
\ba\label{JOcr}
\frac{\partial\widetilde A}{\partial\xi_{\bar I}}=T_{\bar I\bar J}\beta_{\bar J},
\ea
where $\widetilde A$ and $T_{\bar I\bar J}$ are functions of the couplings (except $\theta$) that appear in the Weyl anomaly when all couplings are promoted to space-time dependent functions. Such conditions relate terms of the beta functions of different couplings and different orders in perturbation theory, and can serve as non-trivial consistency checks of multi-loop calculations. For example, eq. \eqref{JOcr} has been employed to resolve an ambiguity in the 4-loop beta function of the strong gauge coupling in the Standard Model \cite{Poole:2019txl}. The very same tool can potentially be used to extract information about the renormalization group evolution in general renormalizable theories like \eqref{LagGen} (see e.g. \cite{Davies:2021mnc}). In that case however one cannot a priori ignore the topological angles. Irrespective of whether they vanish in the bare action, they may be needed as counterterms and therefore generically possess a beta function. Let us see how $\theta$ qualitatively impacts \eqref{JOcr}. 

To start, according to \cite{Keren-Zur:2014sva} the formal structure of \eqref{JOcr} remains unchanged when considering P-violating theories. The absence of an explicit dependence on $\theta$ then implies that its $\bar I=\theta$ component becomes \cite{Keren-Zur:2014sva} 
\ba\label{I=theta}
0=T_{\theta\theta}\beta_\theta+T_{\theta Y}\beta_Y+T_{\theta g}\beta_g+T_{\theta\lambda}\beta_\lambda,
\ea
where $\theta,Y,g,\lambda$ schematically denote the couplings in \eqref{LagGen}. The same considerations developed in Section \ref{sec:renQFT} allow us to identify the tensorial form of $T_{\bar I\bar J}$. At leading order we have
\ba
T_{\theta^{AB}\theta^{CD}}=c_1\frac{g_{AC}^2g_{BD}^2}{(16\pi^2)^3}
\ea
and
\ba
T_{\theta^{AB}Y}dY
&=&c_2\frac{g_{AC}^2g_{BD}^2}{(16\pi^2)^3}{\rm Tr}\left[(Y^\dagger (T^C)^*dY-dY^\dagger (T^A)^*Y)T^D\right]\\\no
&+&c_3\frac{g_{AC}^2g_{BD}^2}{(16\pi^2)^3}{\rm Tr}\left[(Y^\dagger dY-dY^\dagger Y)\left\{T^C,T^D\right\}\right]
\ea
with $c_{1,2,3}$ numerical coefficients. Note that $T_{\theta Y} dY$ must be CP-odd. In extracting its form a key role is played by the fact that CP-odd quantities can here depend on the derivative of the couplings, as opposed to Section \ref{sec:renQFT} when we discussed the form of $\beta_\theta$. Analogously, one can see that the CP-odd structures $T_{\theta g}dg$ and $T_{\theta\lambda}d\lambda$ inevitably arise at a higher perturbative order. Within a 2-loop accuracy the contributions proportional to $\beta_g,\beta_\lambda$ can therefore be neglected and \eqref{I=theta} gives 
\ba
\mu\frac{d\theta^{AB}}{d\mu}
&=&\frac{c_2/c_1}{16\pi^2}{\rm Tr}\left[(Y_a^\dagger (T^A)^*\beta_{Y_a}-\beta_{Y_a}^\dagger (T^A)^*Y_a)T^B\right]\\\no
&+&\frac{c_3/c_1}{16\pi^2}{\rm Tr}\left[(Y_a^\dagger \beta_{Y_a}-\beta_{Y_a}^\dagger Y_a)\left\{T^A,T^B\right\}\right]\\\no
&+&{\cal O}(\hbar^3).
\ea
Plugging in the 1-loop beta function of the Yukawa coupling $Y_a$ we find that the fermionic trace proportional to $c_3$ does not contribute at 2-loops whereas the one proportional to $c_2$ consistently reproduces \eqref{betaTheta} provided $2c_2/c_1=c$. We thus find a fourth independent method for computing the coefficient $c$ of $\beta_\theta$. Establishing its numerical value would require knowing the leading order $T_{\theta\theta}$ and $T_{\theta Y}$, which means performing a 1-loop and again a 3-loop calculation respectively.

The remaining components of \eqref{JOcr} are also quite consequential and deserve to be carefully explored. A naive counting suggests that, in employing the consistency relations of Jack and Osborn with $\bar I=Y$, the 2-loop $\beta_\theta$ identified in this paper might be correlated to structures appearing in the 3-loop $\beta_g$, the 4-loop $\beta_Y$, and the 2-loop $\beta_\lambda$ via the 6-loop contribution in ${\widetilde{A}}$ of order $g^4Y^6$. Furthermore, an inspection of the full system \eqref{JOcr} reveals that the highest perturbative order at which the $g$-$Y$-$\lambda$-$\theta$ beta functions enter in the consistency relations are respectively $5$-$4$-$3$-$2$.


\subsection{(In)Stability of UV solutions of the strong CP problem}
\label{sec:UVstrongCP}

In the Standard Model the first contribution to $\beta_\theta$ is expected to arise at 7-loops \cite{Ellis:1978hq,Khriplovich:1993pf}. This implies that the renormalization group  evolution of $\theta$ is numerically negligible, leaving open the possibility that the absence of CP violation in low energy QCD be the result of some unknown mechanism at very short distances $\sim1/\Lambda$, even close to the Planck scale, that sets $\bar\theta(\Lambda)\approx0$. Among the proposed UV solutions of the strong CP problem we can find heavy axion scenarios or models with approximate CP \cite{Nelson:1983zb,Barr:1984qx} and P \cite{Babu:1989rb,Barr:1991qx}.

The numerical stability of $\theta=0$ is however not a generic property in field theory. We have seen that arbitrary renormalizable theories potentially develop a non-trivial renormalization of the theta angle already at 2-loops. This demonstrates that UV solutions of the strong CP problem cannot work if the effective field theory below $\Lambda$ is completely generic and has unsuppressed Yukawa interactions. This has important phenomenological implications. If no QCD axion is ever to be found, the possibility that some yet unknown UV solution is at play would become increasingly more likely. Tools like those developed here should then be used to identify the class of extensions of the Standard Model that may consistently describe the low energy physics below $\Lambda$.

The radiative instability of UV solutions to the strong CP problem is typically worse than naively expected, actually. First of all, arbitrary extensions of the Standard Model would necessarily feature new mass scales, and it is well-known that threshold corrections can affect the topological angles at tree and loop order. A recent discussion of these corrections in the context of CP-invariant models is given by \cite{Valenti:2021rdu} and in P-invariant models by \cite{deVries:2021pzl} (see \cite{deVries:2018mgf} for a model-independent qualitative assessment). What we want to stress here is another subtle effect: despite the fact that the beta function of $\theta$ is generically at 2-loop order, {\emph{physical}} phases typically renormalize at 1-loop. By {\emph{physical}} phases we mean those special combinations of the renormalized couplings, namely those combinations that are invariant under unitary rotations of the fields, that enter observables. We have seen that $\theta$ is not invariant under anomalous rotations of the fermions, see \eqref{thetaCP}, and therefore the {\emph{physical}} rescaling invariant quantity of interest must take the form 
\ba\label{thetabargen}
\bar\theta=\theta+f(\left\{\xi_i\right\}), 
\ea
where $f$ is a model-dependent function. Here $\bar\theta$ is what we may call the physical, rescaling invariant ``topological" angle. When no $f$ with the required properties exists then no invariant can be built out of $\theta$ and that parameter is unphysical. The physical angle runs according to 
\ba
\beta_{\bar\theta}=\beta_\theta+\frac{\partial f}{\partial\xi_i}\beta_{\xi_i}.
\ea
The symmetry properties of $\beta_{\bar\theta}$ are precisely the same as those of $\beta_{\theta}$. In particular, both must be proportional to the imaginary parts of the invariants $I^{AB}$. Yet, while $\beta_{\theta}$ and $\beta_{\xi_i}$ are polynomial in the couplings, it may happen that $\partial f/\partial\xi_i$ brings inverse powers of the $\xi_i$'s that are not compensated by $\beta_{\xi_i}$. In such a situation $\beta_{\bar\theta}$ develops negative powers of the couplings in front of the invariants $I^{AB}$, which in practice indicates that $\beta_{\bar\theta}$ arises at an order in perturbation theory that is lower compared to $\beta_{\theta}$. We will see explicitly below how this can occur in a concrete model. Importantly, the Standard Model turns out to be special even in this respect because this enhancement cannot appear. Indeed, adopting the standard definition $\bar\theta=\theta-{\rm Im}\,{\rm Tr}\,{\log}[{\cal Y}_u{\cal Y}_d]$, where ${\cal Y}_{u,d}$ denote the Standard Model couplings, we find $\beta_{\bar\theta}=\beta_\theta-{\rm Im}{\rm Tr}\left[{\cal Y}_u^{-1}\beta_{{\cal Y}_u}+{\cal Y}_d^{-1}\beta_{{\cal Y}_d}\right]$. Since by the accidental flavor symmetries $\beta_{\cal Y}=\beta_{\cal Y}' {\cal Y}$, for some polynomial $\beta_{\cal Y}'$, then $\beta_{\bar\theta}$ is made up of the very same polynomial structures of couplings as $\beta_{\theta}$. This fact for example ensures that the estimates of \cite{Ellis:1978hq,Khriplovich:1993pf} are robust.

The faster renormalization group  evolution of the physical theta angles and the presence of important threshold corrections put significant stress on whatever low energy theory emerges from possible UV solutions of the strong CP problem. Rather than carrying out a general analysis of the challenges these scenarios have to face, we find it convenient to discuss a simple representative model.

\subsubsection*{An example}

Consider the following extension of the Standard Model. We introduce two Weyl fermions with charges $\psi\sim{\bf 3}$ and $\psi^c\sim\overline{\bf 3}$ under the QCD group $SU(3)_c$, a singlet scalar $\phi_1$ and a scalar octet $\phi_2\sim{\bf 8}$. These fields are all neutral under the electroweak group. The Lagrangian reads
\ba
{\cal L}&=&{\cal L}_{\rm SM}+\frac{1}{2}(\partial\phi_1)^2+\frac{1}{2}(D\phi_2)^2+\psi^\dagger i{\slashed{D}}\psi+{\psi^c}^\dagger i{\slashed{D}}\psi^c\\\no
&-&[(m+y_1\phi_1)\psi\psi^c+y_2\psi\psi^c\phi_2+{\rm hc}]-V(\phi_1,\phi_2^2,|H|^2).
\ea
In addition to the CP-odd parameters of the Standard Model, this theory features two more phases. On top of that, the rescaling-invariant QCD theta angle contains a new contribution because $\theta_{\rm QCD}$ is now affected by chiral rotations of $\psi,\psi^c$ as well. A natural definition of the phases (the definition of the Jarlskog invariant is the standard one) is given by
\ba\label{phasesEX}
\varphi_y&=&{\rm Arg}\left[(y_1y_2^*)^2\right]
\\\no
\varphi_m&=&{\rm Arg}\left[y_1^2y_2^2[m^*]^4\right]
\\\no
\bar\theta&=&\theta_{\rm QCD}-{\rm Arg}\,{\rm det}[{\cal Y}_u{\cal Y}_d]-{\rm Arg}[m].
\ea
All these CP-odd parameters run already at 1-loop:
\ba\no
\mu\frac{d\varphi_y}{d\mu}
&=&-\frac{4}{16\pi^2}\sin\varphi_y\left[|y_1|^2+\frac{4}{3}|y_2|^2\right]
\\\no
\mu\frac{d\varphi_m}{d\mu}
&=&+\frac{4}{16\pi^2}\left[|y_1|^2\left(\sin\varphi_y-2\sin\frac{\varphi_y+\varphi_m}{2}\right)+\frac{4}{3}|y_2|^2\left(-\sin\varphi_y+2\sin\frac{\varphi_y-\varphi_m}{2}\right)\right]
\\\no
\mu\frac{d\bar\theta}{d\mu}
&=&-\frac{2}{16\pi^2}\left[|y_1|^2\sin\frac{\varphi_y+\varphi_m}{2}+\frac{4}{3}|y_2|^2\sin\frac{-\varphi_y+\varphi_m}{2}\right].
\ea
Note that, as opposed to the Standard Model, this is also true for $\bar\theta$ because the phase of the exotic fermion mass has a non-trivial renormalization group  evolution. As a result $\beta_{\bar\theta}$ arises at 1-loop while our result \eqref{betaTheta} says that 
\ba
\beta_\theta
&=&\frac{c}{(16\pi^2)^2}(+3){\rm Im}\left[(y_1y_2^*)^2\right]+{\cal O}(\hbar^3)\\\no
&=&\frac{3c}{(16\pi^2)^2}|y_1|^2|y_2|^2\sin\varphi_y+{\cal O}(\hbar^3).
\ea
Therefore, in the present model the latter beta function is always subdominant and can be neglected. Indeed it would be enough to have either a small $y_1$ or a small $y_2$ in order to suppress $\beta_\theta$, but this would not stop $\bar\theta$ from running. Similarly, $\sin\varphi_y=0$ would suppress $\beta_\theta$ but not $\beta_{\bar\theta}$ if $\sin\varphi_m\neq0$. Only under the simultaneous conditions $\sin\varphi_y=\sin\varphi_m=0$ the parameter $\bar\theta$ becomes approximately stable under the renormalization group, but this is obvious since in that situation our theory introduces no new CP-odd phases compared to the Standard Model.

It is interesting to verify our earlier claim that the 2-loop $\beta_\theta$ requires at least two scalars with different gauge representations. In the present model 
\ba\label{IAB2model}
I^{AB}_{(2)}=3\,{\rm Im}\left[(y_1y_2^*)^2\right]\delta^{AB}.
\ea
If we replace $\phi_2$ with another scalar singlet (or $\phi_1$ with another scalar in the adjoint) that invariant would not be generated by any diagram. The reason is that in such a scenario the kinetic term of the scalars would possess a $O(2)$ symmetry that rotates $(\phi_1,\phi_2)$. As usual this flavor symmetry is explicitly broken by the interactions, but may be formally resurrected by promoting all the couplings to spurions, as described in Section \ref{sec:renQFT}. The pair $(y_1,y_2)$ would be formally a doublet of $O(2)$ and $\beta_\theta$ should be a singlet. However \eqref{IAB2model} does not meet this requirement. Manifestly $O(2)$ invariants are $y_ay_a$, which is not invariant under axial rotations of the fermions, and $y_ay_a^*$, which is CP-even. There is no combination of the former that is simultaneously CP-odd and invariant under axial rotations, besides of course the one involving $\theta$, which we know has no perturbative effect. Note that the quantity $y_a\epsilon_{ab}y_b=i{\rm Im}[y_1y_2^*]$ is $SO(2)$-symmetric and CP-odd, but in order to have an invariant under $Z_2\subset O(2)$ one would need an even power of it, thus resulting in a CP-even combination. We conclude that $\phi_{1,2}$ cannot be in the same representation if \eqref{IAB2model} is to be generated.

An important point to stress is the following. The definitions \eqref{phasesEX} (and more generally \eqref{thetabargen}) are ambiguous because any combination of invariants is also invariant. One may then object that the claim that the rescaling-invariant topological angle runs at 1-loop is not sufficiently general: maybe there exists a rescaling invariant combination that is also approximately renormalization group  invariant, and one may decide to call it $\bar\theta$. Yet, the ambiguity in defining the physical theta angle arises only from a UV perspective. Its IR definition is basically fixed by experiments. In the Standard Model one usually takes $\theta_{\rm QCD}-{\rm Arg}\,{\rm det}[{\cal Y}_u{\cal Y}_d]$ because at leading order in a perturbative expansion in Yukawas (masses) this is precisely the quantity observed in low energy processes. The situation is similar here. Indeed, because the new fermions $\psi,\psi^c$ are colored, phenomenologically we expect $|m|\gtrsim1$ TeV; at scales below the mass threshold $|m|$ we can thus integrate them out such that our theory reduces to the Standard Model with topological angle given by 
$$
\bar\theta_{\rm SM}=\left\{\theta_{\rm QCD}-{\rm Arg}\,{\rm det}[{\cal Y}_u{\cal Y}_d]-{\rm Arg}[m]\right\}_{\mu=|m|}+{\cal O}(\hbar), 
$$
where the small corrections are due to 1-loop threshold effects. We thus see that the UV ambiguity is resolved: within the accuracy we are working the rescaling-invariant topological angle defined in \eqref{phasesEX} is precisely the one constrained by low energy experiments. The distinction between $\bar\theta_{\rm SM}$ and $\bar\theta$ is 1-loop order, and so the difference in their beta functions starts at 2-loops. Hence the statement that $\bar\theta$ runs already at 1-loop in fact also applies to the physically relevant CP-odd parameter measured in experiments. An UV solution of the strong CP problem that enforces the condition $\bar\theta(\Lambda)=0$ would not be able to explain the absence of CP violation in low energy QCD. To ensure $|\bar\theta_{\rm SM}|\lesssim10^{-10}$ one has to invoke a non-trivial conspiracy among different UV parameters.

We conclude by commenting on what would happen if we set $m=0$. Such an alternative scenario may still be phenomenologically viable provided the exotic fermion gets a mass $y_1v_1$ from the vacuum expectation value of the singlet $\langle\phi_1\rangle=v_1$. This model introduces a unique new phase $\varphi_y$ whereas the QCD topological angle may be defined as $\bar\theta=\theta-{\rm Arg}\,{\rm det}[{\cal Y}_u{\cal Y}_d]-{\rm Arg}\left[y_1^2\right]/2$ as well as in an infinite number of inequivalent ways. Again, the ambiguity in its definition is resolved by observing that the QCD angle at the matching scale reads
\ba\no
\bar\theta_{\rm SM}=\left\{\theta_{\rm QCD}-{\rm Arg}\,{\rm det}[{\cal Y}_u{\cal Y}_d]-{\rm Arg}[y_1v_1]\right\}_{\mu=|y_1v_1|}+{\cal O}(\hbar).
\ea
Again, in order to have a small $|\bar\theta_{\rm SM}|$ there should be a cancellation between the phase of $y_1$, which runs at 1-loop, and $\theta_{\rm QCD}-{\rm Arg}\,{\rm det}[{\cal Y}_u{\cal Y}_d]$, which does not run before 7-loops. This cancellation is not stable unless $\sin\varphi_y=0$, i.e. unless we recover the same amount of CP-violation as in the Standard Model. 

The lesson to be learned is clear. Generic extensions of the Standard Model feature several flavor-invariant CP-odd phases with non-trivial renormalization group  evolution. In such a situation there are no simple UV conditions that ensure $|\bar\theta_{\rm SM}|$ stays small.

\section{Outlook}

It has been known for quite some time that topological angles have measurable consequences. Re-summing the logs associated to their renormalization group evolution might therefore be quantitatively relevant in some case. In this paper we discussed the perturbative running of the $\theta$ angles in general renormalizable theories as well as the techniques necessary to compute it in dimensional regularization. 

We found that $\theta$ can acquire an additive beta function no earlier than 2-loop order. This effect is proportional to a unique combination of the ordinary couplings we called $I^{AB}_{(2)}$. The effect is present only in sufficiently complex theories with (CP-violating) Yukawa interactions involving scalars with more than one irreducible representation of the gauge group. Within the background field method the corresponding divergent diagrams are shown in Fig. \ref{fig:diagrams}.

There are a number of directions in which our work can be extended. An obvious one would be to perform the explicit calculation of the numerical coefficient $c$ of the 2-loop beta function \eqref{betaTheta}. This can be achieved with one, or more, of the four techniques discussed in this paper, namely the background field method \cite{Novikov:1983gd}, the CP-odd spurion trick of \cite{Georgi:1980cn}, or the two new techniques proposed here: via inspection of off-diagonal elements of the anomalous dimension matrix of the CP-odd interactions (see Section \ref{sec:anomalDIM}) or via the Weyl consistency conditions (see Section \ref{sec:consistencyREL}). The calculation may be carried out in any model with a non-vanishing $I^{AB}_{(2)}$, like for example the toy model discussed in Section \ref{sec:UVstrongCP}. In all cases it requires the computation of 3-loop diagrams.

Another direction that needs to be explored is the consistency conditions implied by the local renormalization group approach of Jack and Osborn. In Section \ref{sec:consistencyREL} we have shown that the 2-loop $\beta_\theta$ cannot be ignored compared to the 3-loop $\beta_g$, the 4-loop $\beta_Y$, and the 3-loop $\beta_\lambda$. Yet, in order to draw more quantitative conclusions a systematic generalization of the local renormalization group technique to P-violating theories with anomalous currents, as initiated in \cite{Keren-Zur:2014sva}, as well as a (re)analysis of the beta functions within a consistent scheme for $\gamma_5$, appear necessary.

The phenomenological implications of $\beta_\theta$ also deserve to be further scrutinized. We have seen that the radiative stability of the QCD angle is a peculiar accident of the Standard Model. Quantitatively, a small strong CP phase represents a renormalization-invariant feature of the Standard Model, but certainly not of arbitrary extensions. This observation becomes phenomenologically relevant if the strong CP problem is solved by some mechanism at some high UV cutoff (e.g. models based on approximate P or CP symmetries). In those cases the UV theory would typically predict $\bar\theta(\Lambda)=0$ and the experimental observation $|\bar\theta(1~{\rm GeV})|<10^{-10}$ would translate into a constraint on the renormalization evolution below $\Lambda$. Our considerations can help identify which extensions of the Standard Model at renormalization scales $1~{\rm GeV}<\mu<\Lambda$ are compatible with such a picture. For example, we found that a 2-loop $\beta_\theta$ can only be generated if there are at least two scalars with different representations of the gauge group. Thus, a two-Higgs doublet model would not induce that effect, despite the many additional sources of CP violation. The claim that such an extension features the same radiative stability as the Standard Model is however too naive. We have demonstrated that the {\emph{rescaling-invariant}} angles $\bar\theta$ often run at lower order in perturbation theory than $\theta$ itself and that sizable (finite) effects on $\bar\theta$ usually arise at the mass thresholds that extensions of the Standard Model inevitable possess. UV completions where all these effects are under control are rather rare. Only a very limited number of options would be available if Nature decided to address the strong CP problem in the far UV. The identification of these options represents an interesting open problem.


\section*{Acknowledgments}

We have enjoyed conversations with M. Fael, M. Pospelov, A.E. Thomsen, G. Villadoro. This research was partly supported by the “iniziativa specifica” Physics at the Energy, Intensity, and Astroparticle Frontiers (APINE) of Istituto Nazionale di Fisica Nucleare (INFN), the Italian MIUR under contract 2017FMJFMW (PRIN2017), and the European Union’s Horizon 2020 research and innovation programme under the Marie Sklodowska-Curie grant agreement No 860881-HIDDeN.

\appendix

\section{CP-odd flavor-invariants}
\label{AppIAB}

In this appendix we build the flavor-invariants introduced in Section \ref{sec:renQFT} and identify the CP-odd ones. We approach the problem perturbatively. By counting the powers of $\hbar$ in the flavor-singlet structures one sees that corrections to the 1PI vertex with two external gauge bosons proportional to $g^{2c_g}Y^{2c_Y}\lambda^{c_\lambda}$ correspond to diagrams with $n=c_g+c_Y+c_\lambda$ loops. Yet, the associated contribution to $Z_{\theta,1}$ is a 1-loop factor smaller because the definition of the $\theta$ vertex includes a factor $g^2\hbar/16\pi^2$. This implies that in order to find a correction of $n$-loop size to the beta function, hereafter denoted by $\beta_\theta^{(n)}$, one should calculate an $n+1$-loop diagram. 

On pure dimensional grounds, therefore, corrections to $\beta_\theta$ are expected to be controlled by
\ba
&&\beta_\theta^{(1)}:~~~~~g^4,g^2\lambda,g^2Y^2\\\no
&&\beta_\theta^{(2)}:~~~~~g^6,g^4\lambda,g^2\lambda^2,g^4Y^2,g^2Y^2\lambda,g^2Y^4\\\no
&&\cdots
\ea
In this paper we will content ourselves with 1- and 2-loop size effects, namely $\beta_\theta^{(1,2)}$, though the formalism we adopt can be extended up to arbitrary order. Up to this order it is rather straightforward to argue that only the terms involving the Yukawa couplings have some real chance of being CP-odd, as we now show. 

A general renormalizable gauge theory without scalar quartics and Yukawas always conserves CP (if the topological angles can be neglected, which is the case perturbatively). Hence there cannot be any CP-odd invariant $I^{AB}$ built out of $T^A,S^A$ only. Purely bosonic flavor-invariants cannot work either. They depend on $S^A,\lambda_{abcd}$ and are automatically CP-even. Indeed, $\lambda$ is real whereas $S^A$ are purely imaginary. To build a CP-odd combination we would need an odd number of $S^A$, which cannot be covariant under rotations of the adjoint index because the only invariant tensor at our disposal for contractions is $g_{AB}^2$. This conclusion was expected on account of that the Feynman diagrams we are interested in must be proportional to the Levi-Civita tensor, and therefore fermion traces are strictly necessary to generate them.

We can explicitly demonstrate that even the combinations $g^2\lambda$, $g^2\lambda^2$ and $g^4\lambda$ can be discarded. Recall that a fermion loop is necessary, so such invariant must include traces of the fermion generator. The only $g^2\lambda$ invariant we can have is ${\rm Tr}[T^AT^B]\lambda_{aabb}$ and is manifestly real, i.e. CP-even, since the fermionic trace gives the direct sum of identities in the adjoint index multiplied by (real) fermion Casimirs. Similar considerations apply to the $g^2\lambda^2$ invariants, which are of the form ${\rm Tr}[T^AT^B]\lambda^2$ with all possible contractions of the scalar indices. Whatever contraction is taken the invariant is CP-even. Finally, at order $g^4\lambda$ we can have structures of the form ${\rm Tr}[TTTT]\lambda$, ${\rm Tr}[TT]{\rm Tr}[TT]\lambda$, and ${\rm Tr}[TT]SS\lambda$. The former two are manifestly real because the potential CP-violating contribution would have to come exclusively from the fermionic generators, and we have recalled above that this is not possible. The last one may either contain $[S^MS^N]_{ab}\lambda_{abcc}$ or $[S^MS^N]_{aa}\lambda_{bbcc}$, which are both real. We did not include structures with a single scalar generator because those vanish: the (anti-symmetric) scalar indices in $S^A$ should necessarily be contracted with those (symmetric) of $\lambda$. Similarly, three scalar generators would require the fermionic trace ${\rm Tr} T =0$, which vanishes in the absence of mixed gravity-gauge anomalies.

We conclude that, at least up to the perturbative order considered here, the Yukawa couplings are strictly necessary to build the CP-odd flavor-invariants $I^{AB}$ of the theory \eqref{LagGen}. For convenience it is useful to introduce the basic covariant combinations in which they can appear:
\ba
\Upsilon^{A_1\cdots A_n}_{ab;ij}\equiv Y_{a \, ik}^*(T^{A_1})^*\cdots(T^{A_n})^*Y_{b \, kj}.
\ea
These objects transform under fermion rotations precisely as $T^A$, under scalar rotations as the product of two Yukawas, whereas under gauge boson rotations in an obvious way. We will use this compact notation to write the possible invariants appearing in $\beta_\theta^{(1,2)}$.

\subsubsection*{Absence of $\beta^{(1)}$: 2-loop diagrams}

At lowest order we have a very limited number of flavor-invariant structures that can contribute. They are so few that we can write them explicitly:
\ba\label{1-loopSTRUCTURES}
g^2Y^2:&&{\rm Tr}[T^{(A}T^{B)}\Upsilon_{aa}],~~{\rm Tr}[T^{(A}\Upsilon^{B)}_{aa}],~~{\rm Tr}[T^{(A}T^{B)}]{\rm Tr}[\Upsilon_{aa}],~~S^{(A}_{mn}S^{B)}_{mn}{\rm Tr}[\Upsilon_{aa}],
\ea
Here and in the following ${\rm Tr}[\cdots]$ denotes a trace over the fermionic indices and $(~)$ imply symmetrization. We did not include invariants in which the scalar indices of $S^A_{ab}$ are contracted with those of the Yukawas because thanks to \eqref{SYasYT} these can be written in terms of the first and second invariants of \eqref{1-loopSTRUCTURES}. Furthermore, we did not include structures of the form ${\rm Tr}[\Upsilon^{(AB)}_{aa}]$ because ${\rm Tr}[\Upsilon^{A_1 \cdots A_n}_{ab}] = {\rm Tr}[\Upsilon_{ab} T^{A_n}\cdots T^{A_1}]$ due to the trace transposition property. This relation will also be exploited later in enumerating the invariants of higher order.

We find that none of the invariants in \eqref{1-loopSTRUCTURES} is CP-odd. To see this note that the matrices $\Upsilon_{aa}$, $\Upsilon^A_{aa}$ are hermitian and therefore their trace is real. This immediately tells us that the last two structures are real. Similarly, the structures ${\rm Tr}[T\Upsilon^A]$ are necessarily CP-even because they are the trace of the product of two hermitian matrices, which is real. The first structure in \eqref{1-loopSTRUCTURES} is CP-even for the same reason, because $T^{(A}T^{B) }$ is also hermitian.

\subsubsection*{Terms in $\beta^{(2)}$: 3-loop diagrams}

At the next order, the relevant structures are:
\ba
g^2Y^2\lambda:
&&{\rm Tr}[TT\Upsilon_0]\lambda,~~{\rm Tr}[T\Upsilon_1]\lambda,~~SS{\rm Tr}[\Upsilon_0]\lambda,\\\no
&&{\rm Tr}[TT]{\rm Tr}[\Upsilon_0]\lambda\\\no
g^4Y^2:
&&{\rm Tr}[TT\Upsilon_2],~~{\rm Tr}[TTT\Upsilon_1],~~{\rm Tr}[TTTT\Upsilon_0]\\\no
&&{\rm Tr}[TT]{\rm Tr}[\Upsilon_2],~~SS{\rm Tr}[\Upsilon_2],~~{\rm Tr}[TT]{\rm Tr}[T\Upsilon_1],~~{\rm Tr}[TTT]{\rm Tr}[\Upsilon_1], ~~ SSS{\rm Tr}[\Upsilon_1],  \\\no
&&{\rm Tr}[TTTT]{\rm Tr}[\Upsilon_0], ~~
{\rm Tr}[TT]{\rm Tr}[TT]{\rm Tr}[\Upsilon_0], ~~ SSSS {\rm Tr}[\Upsilon_0]\\\no
g^2Y^4:
&&{\rm Tr}[\Upsilon_2\Upsilon_0],~~{\rm Tr}[\Upsilon_1\Upsilon_1],~~{\rm Tr}[\Upsilon_1\Upsilon_0T],\\\no
&&{\rm Tr}[\Upsilon_2]{\rm Tr}[\Upsilon_0],~~{\rm Tr}[\Upsilon_1]{\rm Tr}[\Upsilon_1],~~{\rm Tr}[TT]{\rm Tr}[\Upsilon_0\Upsilon_0],~~SS{\rm Tr}[\Upsilon_0\Upsilon_0]\\\no
&&{\rm Tr}[\Upsilon_0]{\rm Tr}[\Upsilon_0]{\rm Tr}[TT],~~ SS {\rm Tr}[\Upsilon_0]{\rm Tr}[\Upsilon_0]
\label{invariantStructures}
\ea
where to make our notation more compact the $n$ in the expression $\Upsilon_{n}$ indicates the number of adjoint indices in $\Upsilon$. The indices are left implicit because many contractions are possible, and again invariants where one scalar index must be contracted with a Yukawa have been omitted because of \eqref{SYasYT}. 

\subsubsection*{$g^2Y^2\lambda$ terms}

Most of the $g^2Y^2\lambda$ terms are manifestly real once we take into account that the symmetry of $\lambda$ forces the scalar indices in $[\Upsilon_n]_{ab}$ to be symmetrized or contracted among themselves, i.e. $\Upsilon_{(ab)}$ or $\Upsilon_{aa}$. In either case the resulting $\Upsilon$ tensor is hermitian in the fermion indices. Hence all traces are inevitably real and the invariants CP-even.

\subsubsection*{$g^4Y^2$ terms}
Consider next the $g^4Y^2$ terms. Here the scalar indices are contracted among themselves; it is the possible contractions of the gauge indices that increases the number of independent structures. However all the invariants turn out to be CP-even because of properties of the gauge generators and gauge invariance. Let us show this explicitly by considering the invariants of the form ${\rm Tr}[TT\Upsilon_2]$. These are
\ba
&& g^2 _{CD}{\rm Tr}[T^{(A}T^{B)}\Upsilon^{CD}_{aa}]\\\no
&& g^2 _{CD}{\rm Tr}[T^C T^D \Upsilon ^{(AB)} _{aa}]\
\ea
and
\ba\label{g^4Y^20}
&& g^2 _{CD}{\rm Tr}[T^C T^{(A}\Upsilon^{B)D}_{aa}]\\\no
&& g^2 _{CD}{\rm Tr}[T^C T^{(A|}\Upsilon^{D|B)}_{aa}]\\\no
&& g^2 _{CD}{\rm Tr}[T^{(A|}T^{C} \Upsilon^{|B)D}_{aa}]\\\no
&& g^2 _{CD}{\rm Tr}[T^{(A|}T^{C} \Upsilon^{D|B)}_{aa}]
\ea
where the contraction among adjoint indices is consistently performed with the metric $g^2 _{AB}$, \eqref{LagGen}.
In the first two structures, hermiticity of $T^{(A}T^{B)}$ and $\Upsilon_{aa} ^{(MN)}$ ensures they are all CP-even. The other structures are also CP-even. To see this let us look for a CP-odd version of the first invariant in \eqref{g^4Y^20}, namely
\ba
g^2 _{CD}{\rm Tr}[T^{(A|}T^{C} \Upsilon^{|B)D}_{aa}]-g^2 _{CD}{\rm Tr}[T^{C}T^{(A|} \Upsilon^{D|B)}_{aa}]
\ea
The indices $C,D$ run over all possible adjoint indices. On the other hand, $A,B$ are restricted to a certain non-abelian group or to any two of the abelian factors. Consider first the case $A,B$ refer to a certain non-abelian group. Then $T^{A,B}$ commute with all the $T^{C,D}$ that are associated to the other groups. In addition, $\theta^{AB}\propto\delta^{AB}_{\cal G}$ and hence we should not simply symmetrize but actually sum over $A,B$ as well. As a result the above invariant identically vanishes. Consider next the case in which $A,B$ refer to the abelian groups. Then $T^{A,B}$ commute with $T^{C,D}$ and the expression again identically vanishes. Similar considerations show that all invariants in \eqref{g^4Y^20} are CP-even.

What about invariants of the form ${\rm Tr}[TTT\Upsilon_1]$? Here we have
\ba
&& g_{CD}^2{\rm Tr}[T^{(A|}T^C T^D \Upsilon_{aa} ^{|B)}]\\\no
&& g_{CD}^2{\rm Tr}[T^C T^D T^{(A}\Upsilon^{B)}_{aa}]
\ea
and
\ba\label{g^4Y^22}
&&g_{CD}^2{\rm Tr}[T^{(A|}T^C T^{|B)} \Upsilon_{aa} ^{D}]\\\no
&&g_{CD}^2{\rm Tr}[T^{C}T^{(A|} T^D \Upsilon_{aa} ^{|B)}]\\\no
&&g_{CD}^2{\rm Tr}[T^{(A}T^{B)} T^C \Upsilon_{aa} ^{D}]\\\no
&&g_{CD}^2{\rm Tr}[T^{C}T^{(A} T^{B)} \Upsilon_{aa} ^{D}]
\ea
The expressions with the Casimir $g_{CD}^2T^C T^D=\oplus_{\cal G} C_{\cal G}$ are manifestly real. The first and second in \eqref{g^4Y^22} are real because they are the trace of the product of two hermitian matrices. A potential CP-odd combination with the third structure is
\ba
g_{CD}^2{\rm Tr}[\left\{T^A,T^B\right\}[T^C,\Upsilon_{aa}^D]].
\ea
An identical one is obtained from the last structure in \eqref{g^4Y^22}. Again, these identically vanish when imposing gauge-invariance. Namely, if $A,B$ refer to indices of a non-abelian group, then we should include a sum $\delta_{\cal G}^{AB}$ and find that $\left\{T^A,T^B\right\}$ is proportional to the Casimir $C_{\cal G}$ of that group. Hence it commutes with all $T^C$ and we get $g_{CD}^2{\rm Tr}[C_{\cal G}[T^C,\Upsilon_{aa}^D]]=g_{CD}^2{\rm Tr}[[C_{\cal G},T^C]\Upsilon_{aa}^D]=0$. The same result trivially applies also when $A,B$ refer to the abelian groups.

The arguments just exposed can be applied to all the other invariants appearing in \eqref{invariantStructures}, including the ones with the scalar generators. Therefore our conclusion is that all the $g^4Y^2$ structures are CP-even. 
 
\subsubsection*{$g^2Y^4$ terms}

The structures $g^2Y^4$ are more involved because the scalar indices can be contracted in several different ways. For example ${\rm Tr}[\Upsilon_2\Upsilon_0]$ can be ${\rm Tr}[\Upsilon^{(AB)}_{ab}\Upsilon_{ab}]$, ${\rm Tr}[\Upsilon^{(AB)}_{ab}\Upsilon_{ba}]$, ${\rm Tr}[\Upsilon^{(AB)}_{aa}\Upsilon_{bb}]$. Here it is crucial to observe that $[\Upsilon_{ab}]^\dagger=\Upsilon_{ba}$ if $\Upsilon$ has less than two gauge indices, or even more provided they are symmetrized. This way one finds that ${\rm Tr}[\Upsilon_2\Upsilon_0]$ are all CP-even. Similar considerations apply to ${\rm Tr}[\Upsilon_1\Upsilon_1]$ as well as all the multi-trace expressions in \eqref{invariantStructures}. Including the generator $T$, however, changes the game and leads finally to some interesting candidates. We are thus left with the structure ${\rm Tr}[\Upsilon_1\Upsilon_0T]$. The CP-odd invariants of this form are
\ba\label{invMIO}
&&{\rm Tr}  \left[ \Upsilon^{ A} _{ab} [T^ B , \Upsilon_{ab}] \right] \\\no
&&{\rm Tr}\left[ \Upsilon^{(A} _{ab} [T^{B)}, \Upsilon_{ba}]\right]\\\no
&&{\rm Tr}\left[ \Upsilon^{(A} _{aa} [T^{B)}, \Upsilon_{bb}]\right]
\ea
where the first invariant is already symmetric in the adjoint indices, as follows from the trace transposition property. The same property also shows that the second and third invariants are actually equivalent. Furthermore, using \eqref{SYasYT} it is straightforward to prove that $[\Upsilon_{aa}, T^A]=0$, so the second and third expressions actually vanish. This ensures that the only non-vanishing CP-odd invariant in \eqref{invMIO} is the first one, which is therefore the only possible contribution to the $\beta-$function of $\theta^{AB}$ at 3-loops. This is precisely $I_{(2)}^{AB}$ of eq. \eqref{CPoddInvariantFIN}.

\end{document}